\documentclass[11pt,a4paper]{article}
\usepackage{graphicx}
\usepackage{float}
\usepackage{amsmath, amssymb, accents, amsthm}
\usepackage{centernot}
\usepackage{natbib}
\usepackage{bm}
\usepackage{setspace}
\makeatletter
\def\widebar{\accentset{{\cc@style\underline{\mskip10mu}}}}
\makeatother

\newtheorem{obs}{Observation}
\newtheorem{rem}{Remark}
\newtheorem{exa}{Example}

\begin{document}

\title{From Aggregate Observations to Social Optimum: An Adaptive Pricing Scheme in Heterogeneous Congestion Games\footnote{Remaining errors are my own.}
%\\{\small\it Very Preliminary}
}
\author{Shota Fujishima\footnote{Graduate School of Economics, Hitotsubashi University, 2-1 Naka, Kunitachi, Tokyo 186-8601, Japan. Email: {\tt s.fujishima@r.hit-u.ac.jp}; Website: \texttt{https://sites.google.com/view/sfujishima}}}
\date{April 17, 2025}
\maketitle

\begin{center}
\textbf{Abstract}
\end{center}
\begin{quote}
\hspace{15pt}{\small This study investigates an adaptive pricing scheme aimed at achieving an efficient state in a traffic congestion game characterized by a diverse population of road users. While the planner possesses knowledge of players' preferences, their ability to observe only aggregate states limits the implementation of differentiated taxes. We propose a pricing approach that aligns taxes with the true values of externalities over time, ensuring global stability of the social optimum through replicator dynamics. Our findings suggest that the planner, despite being unable to accurately assess externalities at each moment, can still navigate the economy toward a long-term social optimum by adjusting the disaggregated state based on aggregate observations, while acknowledging the challenges posed by heterogeneous value of time (VOT) among drivers. We also find that a pricing mechanism that incorporates the current externalities for each period, which could be executed by a planner with full access to the disaggregated state, might fail to achieve the global stability of the social optimum under the same dynamics.



}
\end{quote}
\clearpage
\onehalfspacing

\section{Introduction}
In this study, we explore an adaptive pricing scheme that achieves an efficient state in a traffic congestion game involving a diverse population, where the planner is aware of the players' preferences but can only observe aggregate states. Considering that traffic jams include numerous road users, we analyze a population game featuring a continuum of anonymous players.
%\footnote{If players are {\it anonymous}, their allocation does not depend on their identities.} 
In diverse economies, the planner usually has to apply varying taxes on players based on their types, even when they take identical actions. Nevertheless, the anonymity of players compels the planner to implement taxes that are only differentiated by actions. Formally, the social optimum depends on the population distribution over the Cartesian product of action and type sets, which we call a {\it population state}, although the tax policy can depend on only aggregate strategy distributions owing to the planner's informational constraint. Moreover, the planner aims to reach an efficient state represented as a Nash equilibrium; however, this requires each participant to accurately predict the actions of others. In reality, especially within large economies, agents tend to gradually adapt and learn to implement equilibrium strategies. This is the reason we examine an adaptive framework that addresses the stability of the equilibrium results within the context of an agents' learning process governed by evolutionary dynamics.

Sandholm (2002, 05) presents a novel framework to address the implementation problem in congestion games that is practical in large economies. In our congestion game, the planner must internalize the congestion externalities to achieve a social optimum. Sandholm (2002, 05) shows that if the planner can internalize the congestion externalities evaluated at the current state in each period, he can achieve a social optimum in the long run.\footnote{Strictly speaking, we should say "at each instant of time" instead of "in each period" because Sandholm (2002, 05) (and our work) considers continuous-time dynamics.} Specifically, Sandholm considers {\it evolutionary Nash implementation via a price scheme} in which, the planner levies taxes on each action in each period, and the agents gradually learn to play equilibrium strategies according to an evolutionary dynamics. Under Sandholm's price scheme, the planner only needs to internalize the congestion externalities evaluated at the {\it realized} state in each period. In addition, because evolutionary implementation requires the optimum to be globally stable under the evolutionary dynamics, agents learn to play socially optimal strategies from any initial state.

This paper addresses the scenario where the planner is unable to accurately assess the values of externalities {\it even in the current state}. In order to implement Sandholm's pricing scheme, the planner must possess the capability to internalize the externalities at the actual state, which necessitates that players incur uniform costs. Such a condition would be unfavorable, especially in large economies. In our context, the expense of road usage is represented by the time cost, which is calculated as the travel duration multiplied by the agents' value of time (VOT). It would be limiting to presume that a significant portion of drivers have identical VOT. To understand the necessity of homogeneity for Sandholm's model, consider a situation where drivers differ in their VOT, categorized as either high or low. Consequently, given that the extent of externalities experienced by the agents varies based on their VOT, the planner must ascertain the quantity of drivers with high and low VOT to internalize the externalities at the present state. Specifically, the planner must calculate the weighted averages of VOTs for each action, where the weight for a specific type is determined by the share of that type engaging in the action. Nevertheless, since players remain anonymous, the planner can only access the {\it overall} count of drivers. In other words, if agents are diverse, the planner requires insights into the disaggregated state to determine the current values of externalities, despite being able to observe solely the aggregated state.

We take it for granted that the planner is aware of the players' VOTs and the sizes of every population. Therefore, we will not tackle the issue of implementation. Nonetheless, this presumption is not overly limiting in our context since the cost of commuting is closely tied to wages, and planners typically have access to wage distribution data. The scenario we foresee involves the planner categorizing the wage distribution into a set number of bins and designating, for example, the median wage within each bin as the VOTs for those populations.

Because the planner is unable to calculate the values of externalities that arise from the diversity of the players, he cannot effectively navigate the economy toward a social optimum by internalizing the externalities at the existing state in each period. Nevertheless, since the planner's goal is to attain a (static) social optimum in the long-term as outlined by \citet{sandholm2002evolutionary, sandholm2005negative}, it is not essential for externalities to be accurately internalized at the current state in each period; instead, the planner only needs to internalize the externalities in the long-term. Therefore, while we allow agents to play the game under a fixed scheme over time to ensure that the economy reaches a social optimum as \citet{sandholm2002evolutionary, sandholm2005negative} do, we do not require that externalities are consistently internalized along the path to the optimum. In this study, considering that the planner can determine the social optimum using information about players' preferences, we explore a framework where, after assessing the current aggregate state, the planner seeks to {\it adjust} the current disaggregate state towards the optimal one based on a pre-established procedure. The planner aims for his taxes to align with the true values of externalities over the long run. Thus, in contrast to Sandholm's situation, we must clarify how to amend the values of externalities when developing a pricing scheme.

As the planner aims to achieve the social optimum, it is logical to apply the optimality condition, which corresponds with the equilibrium condition under Pigouvian tax policy, when devising a price scheme. In fact, if the planner is successful in enforcing the accurate values of externalities on the agents at equilibrium, this condition must be met. Consequently, we adopt a pricing approach where, during each period, the planner deduces the disaggregated state for every action as if the optimality condition were applicable under his taxes at the observed aggregate state. With this approach, the planner recognizes that his tolls are typically not the accurate values of externalities since the agents do not follow their equilibrium strategies; nevertheless, it is clear that the social optimum is {\it an} equilibrium under this tax policy. However, the fact that a social optimum is an equilibrium does not suffice for the planner, as he guarantees that, as the agents progressively learn to implement these strategies over time, his tolls will eventually mirror the accurate values, thereby ensuring that the optimality condition is fulfilled at equilibrium.

For achieving global equilibrium stability, the introduction of heterogeneity presents a significant technical hurdle to Sandholm's framework. Sandholm's results are derived from the planner's ability to formulate a {\it potential game} \citep{monderer1996PotentialGames} by appropriately adjusting payoff functions via a price scheme. Once a potential game is established, we can analyze the stability of the equilibria within the game by utilizing a corresponding {\it potential function}, as every local maximizer of this potential function represents an asymptotically stable equilibrium when subjected to well-behaved evolutionary dynamics. \citet{sandholm2002evolutionary, sandholm2005negative} illustrates that if the planner charges players the value of congestion externalities calculated at the current state, the game that results is a potential game with the utilitarian social welfare function serving as its potential function. 

One may wish to continue leveraging the characteristics of a potential game as indicated by \citet{sandholm2002evolutionary, sandholm2005negative}; however, the introduction of a heterogeneous cost function renders these properties inapplicable. While potential games are defined by {\it externality symmetry} such that ``the marginal impact of new strategy $j$ users on current strategy $i$ users is the same as the marginal impact of new strategy $i$ users on current strategy $j$ users \citep{sandholm2005negative}," the presence of heterogeneity in the cost function disrupts this symmetry since marginal cost variations typically differ among individuals. Consequently, our game will not support a potential function under any viable pricing strategy. Therefore, rather than adopting the potential game framework, we concentrate on replicator dynamics and identify a Lyapunov function to demonstrate the global stability of the targeted social optimum under our pricing strategy.

If the planner had the capacity to observe the disaggregated state, they could implement Sandholm's scheme where the existing values of externalities are internalized during each period. Nevertheless, this does not automatically enhance the situation in diverse economies. Since the payoff under this approach is derived from the first derivative of the social welfare function, it implies that the social welfare function acts as a potential function. Nonetheless, the social welfare function may exhibit a saddle point, and in such a scenario, achieving global stability of a social optimum is not feasible under well-behaved evolutionary dynamics, including replicator dynamics. This holds true even when a social optimum is globally stable under the replicator dynamics within our toll policy that adheres to the information constraint. This suggests that utilizing disaggregated information can complicate players' incentives, thereby making it challenging to guarantee the achievement of social optimum in the long run.

This study builds upon recent advancements in evolutionary implementation, particularly in the context of heterogeneous economies where planners seek to achieve socially optimal outcomes under limited informational constraints. Among the existing literature, \citet{morimoto2016IEEETrans.Autom.Control} is the closest to our study, both in motivation and methodological approach. Their work investigates multi-agent systems modeled by replicator dynamics and proposes a subsidy-based control mechanism, which, like our pricing scheme, adjusts incentives dynamically based on observed aggregate states rather than individual behaviors. 

Despite these similarities, a key difference between our study and \citet{morimoto2016IEEETrans.Autom.Control} lies in the choice of Lyapunov functions, which naturally leads to different sufficient conditions for global stability under replicator dynamics. While both studies employ Lyapunov methods to analyze stability, the distinct structures of the Lyapunov functions result in different analytical conditions. Our formulation leads to relatively simple sufficient conditions, which help streamline the theoretical analysis and may facilitate broader applicability of the pricing scheme.

Beyond \citet{morimoto2016IEEETrans.Autom.Control}, another related study is \citet{lahkar2021MathematicalSocialSciences}, who analyze evolutionary implementation under the assumption that the planner does not know the agents' preferences but can observe the disaggregated state. Their framework ensures efficiency under evolutionary dynamics when payoffs satisfy certain concavity conditions. In contrast, our study assumes that preferences are known but that the planner can only observe aggregate states, not disaggregated behaviors.

\citet{jinushi2025} examines a different but related setting of information congestion in advertising markets, where multiple agents compete for consumer attention. His study highlights the limitations of traditional Pigouvian taxation in evolutionary settings, demonstrating that static pricing policies may fail to achieve efficiency. While the broader motivation is similar to ours—designing dynamic mechanisms for optimal control—our focus is on traffic congestion externalities rather than information congestion.

The rest of this paper proceeds as follows. In Section 2, we outline the basic framework of the model while defining the evolutionary dynamics and the social optimum. In Section 3, we construct a pricing scheme that relies solely on information regarding aggregate states, and we show that this pricing mechanism secures the global stability of the social optimum under replicator dynamics in scenarios involving two populations and a simple road network. Section 4 serves as the conclusion and reflects on potential subjects for future research. Proofs that are not included in the main text can be found in an Appendix.

\section{Congestion Game with Heterogeneous Population}
Let us consider a heterogeneous extension of \citet[][Example 13.2]{sandholm2015population}. A unit mass of population is divided into a finite number $R$ of groups. The mass of group $r$ is $m_r$, and $\sum_{r\in [R]} m_r =1$, where, for any natural number $N$, we define $[N]$ by $[N] =\{1, 2, ..., N\}$. An infrastructure network is given by a graph $G=(V, E)$ where $V=\{1,2, ... ,n\}$ is the set of nodes and $E$ is the set of edges. Every player commutes from Home to Work. Home is identified as node 1 whereas Work is identified as node $n$. Let $\mathcal{P}$ be the set of (simple) paths from Home to Work. That is, 
\begin{multline}
	\mathcal{P}= \big\{ \{ i_0, i_1, ...., i_t \}\subseteq V: i_0=1, i_t =n,  \\
	(i_k, i_{k+1}) \in E\; \text{for all}\; k=0, ..., n-1\;\text{and}\; \{(i_k, i_{k+1})\}_{k=0}^{t-1}\; \text{are distinct}. \big\}.
\end{multline}
$\mathcal{P}$ is the strategy set of players. Let $L$ and $K$ be the total numbers of edges and paths, respectively. To ease notational burdens, let us assign indices to edges and paths, respectively, so that $E=\{e^1, e^2, ..., e^{L}\}$ and $\mathcal{P}=\{p^1, p^2, ..., p^{K}\}$.

Let $z_{rk}$ be the mass of agents in group $r$ who chooses path $p^k\in \mathcal{P}$.\footnote{We use subscripts $r, s$ to denote groups, subscripts $k, \ell$ to denote actions (or paths), and subscripts $i, j$ to denote links whenever possible.} Let $Z_r= \{ z\in \mathbb{R}_+^{K}:  \sum_{k=1}^K z_{rk} = m_r\}$. Then, $Z\equiv \prod_{r=1}^R Z_r$ is the set of {\it population states}. $X= \{ x\in \mathbb{R}_+^{K}: \exists z\in Z, \forall k\in [K], \sum_{r=1}^R z_{rk} = x_k\}$ is the set of {\it aggregate states}. Let $x(z)$ be the aggregate state associated with population state $z$: i.e., $x_k(z)= \sum_{r=1}^R z_{rk}$ for each $k$.

Let $e(p)$ be the set of edges that comprise path $p$. That is, 
\begin{equation}
	e(p)= \{ (p_i, p_{i+1})\}_{i=0}^{|p|-1},
\end{equation}
where $(p_i, p_{i+1})\in E$ for $i=0, 1, ..., |p|-1$. Let 
\begin{equation}
	\delta_{jk} =  \begin{cases}
		1 & \text{if}\; e^j\in e(p^k), \\
		0 & \text{otherwise}.
		\end{cases}
\end{equation}
$\Delta= [\delta_{jk}]$ is called {\it path-link incidence matrix}. Then, the mass of players who uses edge $e^j\in E$ is
\begin{equation}
	y_j(z) = \sum_{k=1}^K \delta_{jk} x_k(z).
\end{equation}
Let $\theta_r c_j(y_j(z))$ be the group $r$'s cost of using link $e^j\in E$ where $\theta_r >0$ and $c_j$ is an increasing function. The group $r$'s cost of using path $p^k\in \mathcal{P}^r$ is then
\begin{equation}
	C_{rk}(z) = \theta_r \sum_{j=1}^L \delta_{jk} c_j(y_j(z)).
\end{equation}
The payoff from using path $p^k$ for group $r$ is
\begin{equation}
v_{rk} (z) = - C_{rk}(z).
\end{equation}

%A function $W:  Z\to \mathbb{R}$ is a potential function for $v$ if
%\begin{equation}
%\forall r , \forall (k, \ell), \forall z\in Z, \frac{\partial W(z)}{\partial z_k^r} -\frac{\partial W(z)}{\partial z_\ell^r} = v_k^r(z) -v_\ell^r(z).
%\end{equation}

In this paper, we focus on the linear link cost function:
\begin{equation}
c_j(y)=y. 
\end{equation}
Let $\phi_{k\ell} = \sum_{j=1}^{L} \delta_{jk} \delta_{j\ell}$, which is the number of edges that paths $p^k$ and $p^\ell$ share. That is, $\phi_{k\ell} = |e(p^k) \cap e(p^\ell)|$. We then have
\begin{align}
	C_{rk}(z) = \theta_r \sum_{j=1}^{L} \delta_{jk}y_{j}(z) = \theta_r \sum_{j=1}^{L} \delta_{jk} \sum_{\ell =1}^{K} \delta_{j\ell} x_{\ell}(z)= \theta_r  \sum_{\ell=1}^{K} \sum_{j=1}^{L} \delta_{jk} \delta_{j\ell} x_{\ell}(z) .
%	\\
%	&= \theta^r  \sum_{\ell=1}^{K} \phi_{k\ell} \sum_{s\in \mathcal{R}^\ell} z_\ell^s \\
%		&=\theta^r   \sum_{\ell=1}^K\sum_{s\in \mathcal{R}^\ell}\phi_{k\ell}  z_\ell^s.
\end{align}
Let $v_r=(v_{rk})_{k\in [K]}'$ and $\Phi= [\phi_{k\ell}]$. Then,
\begin{equation}
\underbrace{v_r}_{K \times 1} = - \theta_r \underbrace{\Phi}_{K \times K} \underbrace{x}_{K\times 1}.
\end{equation}

Note that $x(z)=(I_K, I_K, ..., I_K)z= (\bm 1_R'\otimes I_K)z$ where $I_K$ is the $K$-dimensional identity matrix and $\bm 1_R$ is the $R$-dimensional vector of ones. Hence,
\begin{equation}
v = - (\theta \otimes \Phi) (\bm 1_R'\otimes I_K) z = - [(\theta \bm 1_R')\otimes \Phi]z= -A z,
\end{equation}
where $v= (v_1, ..., v_R)'$, $\theta=(\theta_1, ..., \theta_R)'$, and $A= [(\theta \bm 1_R')\otimes \Phi]$. We define a population game using $v$ (i.e., a population game featuring a unit mass of players, the action set $\mathcal{P}$, and the payoff vector $v$).

To characterize Nash equilibrium, we represent $A$ as the following segmented matrix:
\begin{equation}
A = \begin{pmatrix}
A_1\\
\vdots \\
A_{R}
\end{pmatrix},
\end{equation}
where $A_{r} = (\theta_r \bm 1_R')\otimes \Phi$. Additionally, let $z_r = \{z_{rk} \}_{k\in [K]}\in Z^r$ for each $r\in [R]$. Then, $\bar{z}\in Z$ is a {\it Nash equilibrium} of $v$ if
\begin{equation}
\forall r\in [R], \forall z\in Z,  - (z_r- \bar{z}_r)' A_r\bar{z} \le 0.
\end{equation}

For the stability of Nash equilibria, we consider the {\it replicator dynamics}: for each $r\in [R]$,
\begin{align}
\dot{z}_r &=- \text{diag}(z_r)( A_r z - \frac{1}{m_r}\bm 1_K z_r' A_r z) \\
&= - (\text{diag}(z_r)  - \frac{1}{m_r}z_r z_r') A_r z . \label{eq:replicator}
\end{align}

\subsection{Social welfare} 
We define the social welfare by the total sum of payoffs. That is, we consider the following utilitarian social welfare function:
\begin{equation}
SW(z) = - z'Az.
\end{equation}

For single-population congestion games, the utilitarian social welfare function is strictly concave. To see whether this is true here, we examine whether $B= \frac{1}{2}(A+A')$ projected to the tangent space of $Z$, which is denoted by $TZ$, is positive definite. To this end, we consider the $K\times (K-1)$ matrix $Q$ defined by
\begin{equation} \label{Q}
	Q= \begin{pmatrix}
		1 &  \\
		& 1 & \\
		&& \ddots &\\
		&& & 1 \\
		-1 &-1 &\cdots & -1
		\end{pmatrix}.
\end{equation}
$Q$ forms a basis of $TZ_r $. That is, $TZ_r =\{Qz: z\in \mathbb{R}^{K-1}\}$. Then,
\begin{equation}
	\underset{RK\times R(K-1)}{\mathcal{Q}} = \begin{pmatrix} Q & \\ &\ddots \\ && Q\end{pmatrix}
\end{equation}
forms a basis of $TZ$. Observe that $\Phi =\Delta'\Delta$ and hence, $\Phi$ is symmetric. Therefore, $B= \frac{1}{2}(\theta \bm 1_R' +\bm 1_R \theta')\otimes \Phi$. We then project $B$ to $TZ$ by
\begin{equation}
	\mathcal{Q}' B \mathcal{Q} = \frac{1}{2}(\theta \bm 1_R' +\bm 1_R \theta') \otimes Q'\Phi Q.
\end{equation}

Let $\lambda$ [resp. $\tau$] be an eigenvalue of $\frac{1}{2}(\theta \bm 1_R' +\bm 1_R \theta')$ [resp. $Q'\Phi Q$]. Then, $\lambda \tau$ is an eigenvalue of $\mathcal{Q}'B\mathcal{Q}$ \citep[][Theorem 21.11.1]{harville2008matrix}. The eigenvalues of $\theta \bm 1_R' +\bm 1_R \theta'$ are
\begin{equation}
	0, \, \sum_{r=1}^R \theta_r +  \sqrt{R}\sqrt{\sum_{r=1}^R(\theta_r)^2}, \, \sum_{r=1}^R \theta_r - \sqrt{R} \sqrt{\sum_{r=1}^R(\theta_r)^2},
\end{equation}
where the multiplicity of the eigenvalue zero is $R-2$. On the other hand, the $(k, k)$-the element of $Q'\Phi Q$ is
\begin{equation}
	\phi_{kk} - \phi_{kK} + \phi_{KK} - \phi_{Kk}.
\end{equation}
Observe that $\phi_{kk} = |e(p^k)| > |e(p^k) \cap e(p^K)|= \phi_{kK}$. Similarly, $\phi_{KK} >\phi_{Kk}$. Therefore, $\sum_i \tau_i = \text{tr}(Q'\Phi Q)  >0$, which implies that $Q'\Phi Q$ has at least one nonzero (in particular, positive) eigenvalue. On the other hand, $\theta \bm 1_R' +\bm 1_R \theta'$ has both positive and negative eigenvalues. Hence, $\mathcal{Q}'B\mathcal{Q}$ also has both positive and negative eigenvalues. 

\begin{obs}\label{obs:sw}
	The social welfare function $SW(z)=-z'Az$ is neither concave nor convex on $Z$.
\end{obs}

\subsection{Adaptive Pigouvian policy}
Because $\frac{\partial SW(z)}{\partial z'}= (A+ A')z$, by imposing a toll of 
\begin{equation}
p(z)= A'z 
\end{equation}
on the agents, the resultant game achieves a payoff vector of $v^p = (A+A')z$, which qualifies as a potential game with the social welfare function acting as its potential function. Whereas the Pigouvian policy sets the values of externalities based on the optimal state, $p$ sets the values based on the {\it current} state of the system. This leads us to refer to $p$ as the {\it adaptive Pigouvian policy}. It is important to note that the planner {\it cannot} execute this policy without the ability to observe the population state $z\in Z$. Since the social welfare function serves as a potential function for $v^p$, the equilibria of $v^p$ correspond to the KKT points of the maximization problem $\max_{z\in Z}\, SW(z)$ \citep[See, e.g., ][]{sandholm2015population}. Therefore, while social optima are equilibria of $v^p$, Observation \ref{obs:sw} indicates that $v^p$ might also possess additional equilibria that do not represent optimal outcomes. Specifically, we have noted that the (projected) Hessian of $SW$ contains both positive and negative eigenvalues, indicating that a non-optimal equilibrium can exist within a stable subspace. Consequently, the global stability of social optima cannot be assured. Example \ref{exa:two_pop_identity} in the following section illustrates a scenario where this situation indeed occurs as shown in Figure \ref{fig:adaptive_Pigouvian}.

\section{Aggregate-based congestion pricing}
We consider aggregate-based congestion pricing that utilizes only the accessible information (i.e., aggregate state or overall traffic volume on each link) instead of the population state or breakdown of traffic volume by individual values of time. Although the planner can only observe aggregate states, it is presumed that the planner is aware of the time values (i.e., $\theta$) and their distribution within the population (i.e., $\{m_r\}_{r\in [R]}$), and thus, this information is utilized. Specifically, considering that the planner can determine the social optimum $z^\ast \in Z$, the planner deduces that the current population state is $z_{rk}=\psi_{rk}x_k$ for each $(r, k)\in [R]\times [K]$ given the aggregate state $x\in X$, where
\begin{equation}
\psi_{rk}= \frac{z_{rk}^\ast}{x_k^\ast}.
\end{equation}
In matrix form, 
\begin{equation}
z= \psi(x) \equiv \Psi x, 
\end{equation}
where $\Psi$ is the $(N\times K)$-matrix given by
\begin{equation}
\Psi=
\begin{pmatrix}
\text{diag}(\{\psi_{1k}\}_{k\in [K]})\\
\vdots \\
\text{diag}(\{\psi_{Rk}\}_{k\in [K]})
\end{pmatrix} .
\end{equation}
$\text{diag}(x)$ is the diagonal matrix having vector $x$ as its diagonal elements.

Then, we consider the following toll policy:
\begin{equation}
\tau(x) \equiv p(\psi (x)) = A' \Psi x.    
\end{equation}
Because $x = (\bm 1_R'\otimes I_K ) z$, the payoff vector under this toll policy is
\begin{align}
v^\tau &=A z+ A'\Psi  (\bm 1_R'\otimes I_K )z \\
&= Az + \begin{pmatrix}
\theta_1 \Phi & \cdots & \theta_R\Phi \\
\vdots & \ddots & \vdots \\
\theta_1 \Phi & \cdots & \theta_R \Phi
\end{pmatrix} \begin{pmatrix} \Psi_1 \\ \vdots \\ \Psi_R\end{pmatrix} \begin{pmatrix} I_K &\cdots &I_K\end{pmatrix} z \\
&= \underbrace{[A+ J_R\otimes \sum_r \theta_r\Phi \Psi_r] }_{\equiv A^\tau} z ,
\end{align}
where $J_R$ represents the $(R, R)$-matrix filled with ones. Let
\begin{equation}
A^\tau_r = \bm 1_K' \otimes [ \theta_r \Phi + \sum_s \theta_s\Phi \Psi_s]. \label{eq:agg_based_pricing_pop}
\end{equation}
Then, the payoff vector of population $r$ is $v_r^\tau = A_r^\tau z$. Note that \eqref{eq:agg_based_pricing_pop} indicates that agents following a route incur the same fee no matter their VOT, making the toll policy viable.

\subsection{Two populations and a network with two independent paths}
\begin{figure}[H]
\centering
\includegraphics[scale=.3]{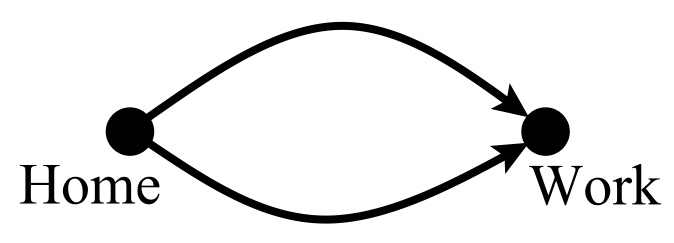}
\caption{Network with two independent paths}
\label{fig:network_two_paths}
\end{figure}

We consider cases where there are two populations, and Home and Work are connected by two independent paths as in Figure \ref{fig:network_two_paths}. 
That is,  we assume that $K=R=2$ and $\Phi$ is a two-dimensional identity matrix. In this case, we have
\begin{equation}
A = \begin{pmatrix}
\theta_1& &\theta_1 \\
&\theta_1 && \theta_1  \\
\theta_2& &\theta_2 \\
&\theta_2 && \theta_2  \\
\end{pmatrix}.
\end{equation}
By the following observation, this game has a continuum of equilibria.
\begin{obs}\label{v_congestion}
Suppose $K=R=2$ and $\Phi= I_K$. The set of Nash equilibria of $v$ is $\{z\in Z: z_{11}+z_{21} = \frac{1}{2}\}$.
\end{obs}
\begin{proof}
See Appendix.
\end{proof}

In Observation \ref{obs:sw}, it has been noted that the social welfare function exhibits neither concavity nor convexity. Let us take a closer look at the function's shape in the case we are considering here. When $K=2$, $Q=(1, -1)'$. Hence,
\begin{equation}
\mathcal{Q} = \begin{pmatrix} 1& \\ -1& \\ & 1\\ & -1 \end{pmatrix},
\end{equation}
Then, we have
\begin{align}
\mathcal{Q}'B\mathcal{Q} &= \frac{1}{2}(\theta \bm 1_R'+ \bm 1_R \theta') \otimes Q'\Phi Q\\
&= \begin{pmatrix} \theta_1 & \frac{\theta_1+\theta_2}{2} \\\frac{\theta_1+\theta_2}{2} &\theta_2 \end{pmatrix} \otimes Q'Q \\
&= \begin{pmatrix} 2\theta_1 & \theta_1+\theta_2 \\ \theta_1+\theta_2 & 2\theta_2 \end{pmatrix} \quad \because Q'Q =2.
\end{align}
Eigenvalues of $\mathcal{Q}'B\mathcal{Q}$ are
\begin{equation}
\theta_1+\theta_2\pm \sqrt{2}\sqrt{(\theta_1)^2+(\theta_2)^2}.
\end{equation}
If $\theta_1\neq \theta_2$, $\theta_1+\theta_2- \sqrt{2}\sqrt{(\theta_1)^2+(\theta_2)^2}<0$. Hence, we have the following.
\begin{obs}
Suppose $K=R=2$ and $\Phi= I_K$. Then, $SW$ has a saddle point on $Z$.
\end{obs}

\begin{exa}[$m_1=m_2=\frac{1}{2}, \theta_1=\frac{1}{2}, \theta_2=\frac{3}{2}$]\normalfont\label{exa:two_pop_identity}
The social optimum is reached at $(z_{11}, z_{21})=(0, \frac{5}{12})$ and $(z_{11}, z_{21})=(\frac{1}{2}, \frac{1}{12})$, which are indicated as red bullets in Figure \ref{fig: eqm_so}. Additionally, $(z_1^1, z_1^2)=(\frac{1}{4}, \frac{1}{4})$ serves as the saddle point of $SW$, shown as the blue bullet in the figure. The arrows depicted in the figure represent vector fields produced by replicator dynamics.
\begin{figure}[H]
\centering
\includegraphics[scale=1]{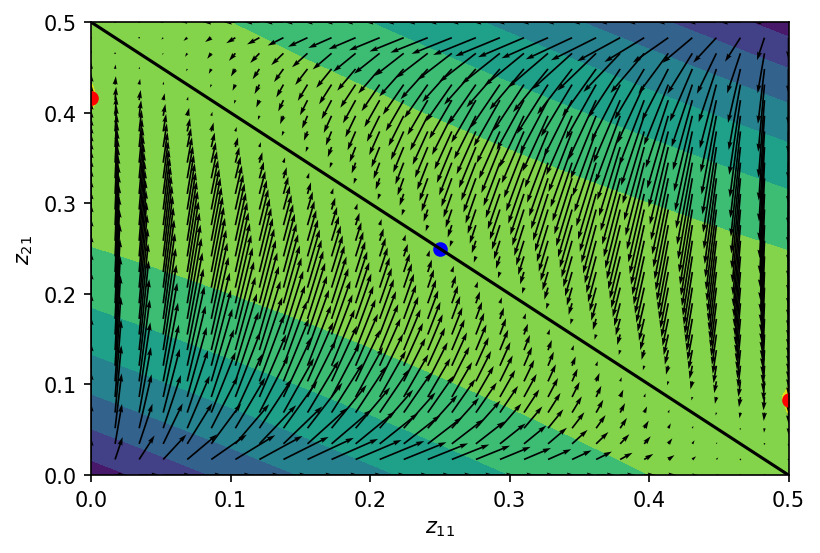}
\caption{The contour graph of $SW(z_{11}, m_1-z_{11}, z_{21}, m_2-z_{21})$ and the equilibria of $v$ are shown for Example \ref{exa:two_pop_identity}, where the function exhibits higher values in the lighter areas and lower values in the darker areas. The black line sloping downwards illustrates the equilibria of $v$. The red dots signify social optima, while the blue dot marks the saddle point of $SW$. The arrows depict vector fields produced by replicator dynamics.}
\label{fig: eqm_so}
\end{figure}
\end{exa}

\begin{figure}[H]
\centering
\includegraphics{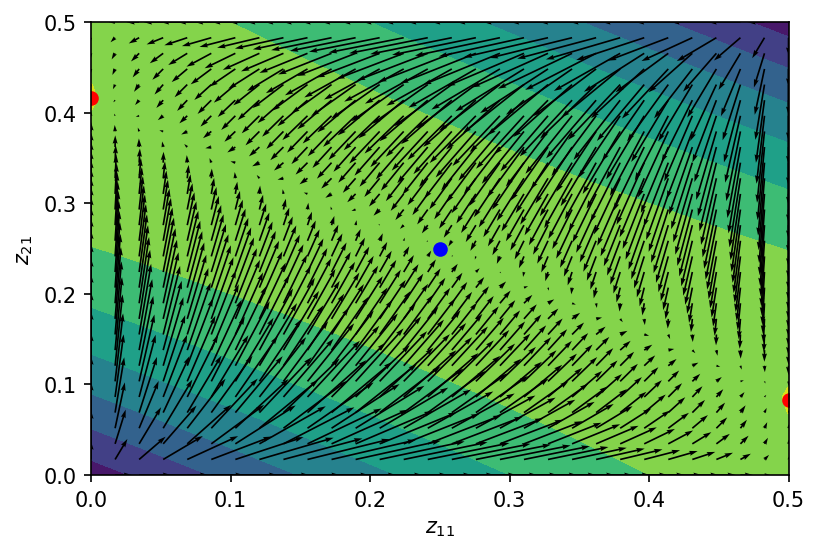}
\caption{The contour graph of $SW(z_{11}, m_1-z_{11}, z_{21}, m_2-z_{21})$ and the equilibria of $v^p$ are shown for Example \ref{exa:two_pop_identity}, where the function exhibits higher values in the lighter areas and lower values in the darker areas.  The arrows depict vector fields produced by replicator dynamics. The red dots represent locally stable equilibria of $v^p$ that are optimal, while the blue dot signifies another equilibrium of $v^p$ that possesses a stable subspace.}
\label{fig:adaptive_Pigouvian}
\end{figure}

Figure \ref{fig:adaptive_Pigouvian} illustrates vector fields produced by replicator dynamics under the adaptive Pigouvian policy $p$. As we have discussed, the equilibria of $v^p$ align with the KKT points of $\max_{z\in Z}\, SW(z)$. Consequently, the equilibria are represented by the three bullets in Figure \ref{fig:adaptive_Pigouvian}. Among these, the red bullets signify social optima that exhibit local stability. However, the blue bullet, which represents the saddle point of $SW$, also possesses a stable subspace. Therefore, the planner cannot assure that the economy will reach a social optimum.

Finally, we will examine the stability of the social optimum under the aggregate-based toll policy $\tau$. To this end, let $B^\tau = \frac{1}{2}(A^\tau + (A^\tau)')$. Projecting this onto $TZ$ yields
\begin{align}
&\mathcal{Q}'B^\tau \mathcal{Q} \\
&= \mathcal{Q}'\begin{pmatrix}
\theta_1 + \sum_r \theta_r\psi_{r1} & 0 &\frac{\theta_1+\theta_2}{2} +   \sum_r \theta_r\psi_{r1} &  \\
0 & \theta_1 + \sum_r \theta_r\psi_{r2} & 0 &\frac{\theta_1+\theta_2}{2} +   \sum_r \theta_r\psi_{r2}  \\
\frac{\theta_1+\theta_2}{2}+ \sum_r \theta_r\psi_{r1}& 0 & \theta_2+ \sum_r \theta_r\psi_{r1} & 0 \\
0 & \frac{\theta_1+\theta_2}{2}+ \sum_r \theta_r\psi_{r2} & 0 & \theta_2+  \sum_r \theta_r\psi_{r2}
\end{pmatrix}\mathcal{Q} \\
&= \begin{pmatrix}
2\theta_1 + \sum_r \theta_r (\psi_{r1}+\psi_{r2}) & \theta_1 +\theta_2+ \sum_r \theta_r (\psi_{r1}+\psi_{r2})\\
\theta_1 + \theta_2+ \sum_r \theta_r (\psi_{r1}+\psi_{r2}) & 2\theta_2 + \sum_r \theta_r (\psi_{r1}+\psi_{r2})
\end{pmatrix}.
\end{align}
The eigenvalues of this matrix are
\begin{equation}
\theta_1+\theta_2 +\delta \pm \sqrt{2(\theta_1^2+\theta_2^2) +2 \delta (\theta_1+\theta_2) + \delta^2},
\end{equation}
where $\delta = \sum_r \theta_r (\psi_{r1}+\psi_{r2})$. Consequently, $\mathcal{Q}'B^\tau\mathcal{Q}$ contains both positive and negative eigenvalues, indicating that $z^\ast$ does not qualify as a {\it Taylor evolutionarily stable state (ESS)}, which is a sufficient condition for stability under replicator dynamics \citep[See, e.g., ][Section 5]{sandholm2010LocalStabilityEvolutionary}.

\begin{figure}[H]
\centering
\includegraphics{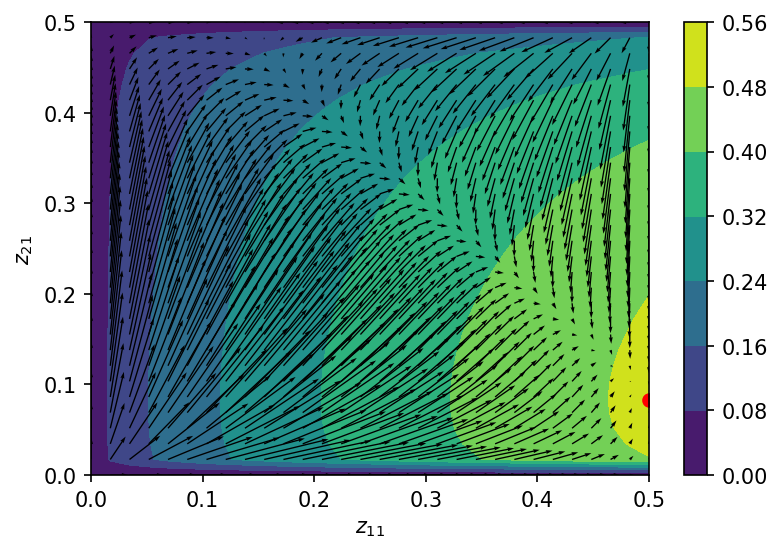}
\caption{The contour graph of Lyapunov function $V(z_{11},m_1-z_{11},z_{21},m_2-z_{21})$ and the equilibrium of $v^\tau$ are shown for Example \ref{exa:two_pop_identity}. The arrows depict vector fields produced by replicator dynamics. The red dot marks the unique equilibrium $v^\tau$, which is globally stable under replicator dynamics.}
\label{fig:agg_based_pricing}
\end{figure}

However, Figure \ref{fig:agg_based_pricing}, which depicts vector fields generated by replicator dynamics under the aggregate-based toll policy $\tau$ targeting $z^\ast = (\frac{1}{2}, 0, \frac{1}{12}, \frac{5}{12})$, suggests that $z^\ast$ is globally stable, and Observation \ref{obs:agg_pricing_gs} confirms this outcome. In Observation \ref{obs:agg_pricing_gs}, global stability is established by demonstrating that the following function acts as a Lyapunov function:
\begin{equation}\label{eq:lyapunov}
%V(z) = \sum_{(r, k): z_{rk}^\ast >0}q_r z_{rk}^\ast \ln \frac{z_{rk}^\ast}{z_{rk}},
V(z) = \prod_{r\in [R]}\prod_{k\in [K]} (z_{rk})^{q_rz_{rk}^\ast},
\end{equation}
where $q_1 =1, q_2 = \frac{2\theta_1+ \sum_t \theta_t  (\psi_{t1}+\psi_{t2})}{2\theta_2+\sum_t \theta_t  (\psi_{t1}+\psi_{t2})}$.

\begin{obs}\label{obs:agg_pricing_gs}
Suppose $K=R=2$ and $\Phi=I_K$. Then, the aggregate-based toll policy $\tau$ ensures that the targeted social optimum is globally stable under replicator dynamics.
\end{obs}
\begin{proof}
See Appendix.
\end{proof}

\begin{rem}\normalfont
In multipopulation games, there exists a less stringent stability concept referred to as {\it Cressman ESS} \citep{cressman2001EvolutionaryStabilityConcepts, garay2000StrictESSNspecies}. In this study, we explicitly demonstrate stability under replicator dynamics by constructing a Lyapunov function, thus the status of $z^\ast$ as a Cressman ESS remains to be verified.
\end{rem}

\section{Conclusion}
To be written.

\section*{Appendix}
\begin{proof}[Proof of Observation \ref{v_congestion}]
The equilibrium conditions are
\begin{gather}
z_{11} > 0\Rightarrow z_{11}+z_{21} \le \frac{1}{2}, \label{eqm1_a}\\
z_{11} < m_1 \Rightarrow z_{11}+z_{21} \ge \frac{1}{2}, \label{eqm2_a} \\
z_{21} > 0\Rightarrow z_{11}+z_{21} \le \frac{1}{2}, \label{eqm3_a}\\
z_{21} < m_2 \Rightarrow z_{11}+z_{21} \ge \frac{1}{2}. \label{eqm4_a}
\end{gather}By these conditions, population state $z\in Z$ such that $z_{11}+z_{21}=\frac{1}{2}$ is obviously a Nash equilibrium. We then show that $z_{11}+z_{21}=\frac{1}{2}$ at any equilibrium. Suppose $z_{11}+ z_{21} <\frac{1}{2}$. Then, by (contrapositions of) \eqref{eqm2_a} and \eqref{eqm4_a}, $z_{11}=m_1$ and $z_{21}=m_2$, and hence $z_{11}+z_{21} = 1$, which is a contradiction. Similarly, we cannot have $z_{11}+z_{21} > \frac{1}{2}$.  
\end{proof}

\begin{proof}[Proof of Observation \ref{obs:agg_pricing_gs}]
Let us consider the function given by \eqref{eq:lyapunov}. We have
\begin{align}
\Dot{V}(z) &= V(z)\sum_{r\in [R]}\sum_{k\in [K]} q_r z_{rk}^\ast \frac{\dot{z}_{rk}}{z_{rk}} \\
&= - V(z)\sum_{r\in [R]}q_r (z_{r}^\ast - z_r)' A_r^\tau z. \quad \because \eqref{eq:replicator} 
\end{align}
Note that 
\begin{align}
A_r^\tau &=( \bm 1_K' \otimes [ \theta_r I_K + \sum_s \theta_s \Psi_s])z \\
&=\underbrace{\left[\theta_r I_K + \begin{pmatrix} \sum_s \theta_s \psi_{s1} & 0 \\
0 &  \sum_s \theta_s \psi_{s2} \\
\end{pmatrix}\right]}_{\equiv B_r^\tau} x.
\end{align}
Hence,
\begin{equation}
\Dot{V}(z) = -V(z)[(z_1^\ast -z_1)' B_1^\tau x+q_2 (z_2^\ast -z_2)' B_2^\tau x]
\end{equation}
Because $z^\ast$ is an equilibrium of $v^\ast$,  $-(z_r- z_r^\ast)'B_r^\tau x^\ast \le 0$. Hence,
\begin{align}
&- (z_r^\ast -z_r) B_r^\tau x  \\
&\ge (z_r^\ast- z_r)'B_r^\tau x^\ast - (z_r^\ast -r_2) B_r^\tau x  \\
&=(z_r^\ast -z_r)'B_r^\tau (x^\ast -x).
\end{align}
%On the other hand, $z_{11}^\ast =m_1, z_{12}^\ast =0$. We assume that the equilibrium is quasi-strict, hence $(A_1x^\ast)_1 < (A_1x^\ast)_2$.\footnote{Note: the payoff is $-(A_rx)_k$.} Hence,  $ (z_1^\ast -z_1)'A_1x^\ast=m_1 (A_1 x^\ast)_1 - z_1' A_1x^\ast = (m_1-z_{11})(A_1x^\ast)_1 - z_{12}(A_1x^\ast)_2 < (m_1-z_{11})(A_1x^\ast)_1 - z_{12}(A_1x^\ast)_1=0$. Hence,
Therefore,
\begin{align}
&- (z_1^\ast -z_1)' B_1^\tau x- q_2(z_2^\ast -z_2)' B_2^\tau x \\
&\ge (z_1^\ast -z_1)' B_1^\tau(x^\ast-x)  +q_2(z_2^\ast -z_2)'B_2^\tau (x^\ast -x)\\
&= (z_1^\ast -z_1)' B_1^\tau(z_1^\ast -z_1) + (z_1^\ast -z_1)' B_1^\tau(z_2^\ast -z_2) \\
&\quad +q_2 (z_2^\ast -z_2)'B_2^\tau(z_2^\ast -z_2) + q_2(z_2^\ast -z_2)'B_2^\tau(z_1^\ast -z_1),
\end{align}
where the last equality follows because $x^\ast - x = z_2^\ast  - z_2 + z_1^\ast - z_1$.

Let $\xi_1 = z_1^\ast -z_1$ and $\xi_2 = z_2^\ast- z_2$. Because $\xi_r$ belongs to the tangent space of $\Delta_r$ for each $r$, we describe them by $\xi_1=(\alpha, -\alpha)$ and $\xi_2=(\beta, -\beta)$, respectively.\footnote{In Example \ref{exa:two_pop_identity}, $\xi_1 = z_1^\ast -z_1 = (m_1-z_{11}, -z_{12})'$, and hence we must have $\alpha\ge 0$.} Then,
\begin{gather}
\xi_1'B_1^\tau\xi_1 = 2\theta_1\alpha^2 + \delta \alpha^2 , \\
q_2\xi_2'B_2^\tau\xi_2 = 2q_2\theta_2 \beta^2 + q_2\delta \beta^2  , \\
\xi_1' B_1^\tau \xi_2 + q_2\xi_2'B_2^\tau\xi_1 =2 (\theta_1+q_2\theta_2)\alpha\beta + (1+q_2)\delta \alpha \beta,
\end{gather}
where $\delta = \sum_t \theta_t  (\psi_{t1}+\psi_{t2})$. Solving
\begin{equation}
2\sqrt{(2\theta_1+\delta)(2q_2\theta_2+\delta)} = 2 (\theta_1+q_2\theta_2) + (1+q_2)\delta
\end{equation}
for $q_2$ yields $q_2 = \frac{2\theta_1+\delta}{2\theta_2+\delta}$. Therefore,
\begin{equation}
\Dot{V}(z) \ge V(z) (\sqrt{2\theta_1+\delta}\alpha + \sqrt{2q_2\theta_2+\delta}\beta)^2 \ge 0.
\end{equation}
\end{proof}

\bibliography{references}
\end{document}